\documentclass[onecolumn,showpacs]{revtex4}

\usepackage{graphicx}
\usepackage{dcolumn}
\usepackage{bm}

\input epsf

\begin{document}

\title{\Large Junction Conditions and Consequences of Quasi-Spherical Space-Time with Electro-Magnetic Field and Vaidya Matric}

\author{\bf Soma Nath$^1$, Ujjal
Debnath$^2$\footnote{ujjaldebnath@yahoo.com}and ~Subenoy
Chakraborty$^1$\footnote{subenoyc@yahoo.co.in}}

\affiliation{$^1$Department of Mathematics, Jadavpur University,
Calcutta-32, India.\\ $^2$Department of Mathematics, Bengal
Engineering and Science University, Shibpur, Howrah-711 103,
India.\\}

\date{\today}

\begin{abstract}
In this work the junction conditions between the exterior
Reissner-Nordstrom-Vaidya space-time with the interior
quasi-spherical Szekeres space-time have been studied for
analyzing gravitational collapse in the presence of a
magneto-hydrodynamic fluid undergoing dissipation in the form of
heat flow. We have discussed about the apparent horizon and have
evaluated the time difference between the formation of apparent
horizon and central singularity.
\end{abstract}


\maketitle

\section{\normalsize\bf{Introduction}}
In Einstein gravity, gravitational collapse is an important
problem for astrophysical objects. Usually, the formation of
compact stellar objects such as white dawrf and neutron star are
preceded by a period of collapse. Hence, for astrophysical
collapse, it is necessary to describe the appropriate geometry of
interior and exterior regions and to determine proper junction
conditions which allow the matching of these regions.\\

The study of gravitational collapse was started long ago in 1939
by Oppenheimer and Snyder [1] to solve an idealized problem of
gravitational collapse preceded by assuming a spherical symmetric
distribution, adiabatic flow and the equation of state in the
form of dust. Misner and Sharp [2] have introduced pressure
gradient terms in the equation of motion. Outgoing radiation of
the collapsing body has been considered by Vaidya [3], Santos [4]
and collaborators [5 - 10] and the correspondence to adiabatic
distribution of matter has been considered by Misner [11] and
collaborators [12, 13]. Models of radiating collapse with viscous
fluid were presented by Lake [14] and Santos [4]. Ghosh and
Deshkar [15] have considered collapse of a radiating star with a
plane symmetric boundary (which has a close resemblance with
spherical symmetry [16]) and have concluded with some general
remarks.\\

Most of the studies has considered a collapsing star whose
interior geometry is spherical. But in the real astrophysical
situation, the geometry of the interior of a star may be
quasi-spherical in form. Previously we have discussed [17]
junction condition between interior Szekeres model and exterior
plane symmetric Vaidya space-time. Oliveira et al [9] have
proposed to study the junction conditions of a collapsing
non-adiabatic charged body producing radiation. The interior
space-time ~$V^{-}$~ was modeled by a magneto-hydrodynamic fluid
undergoing dissipation in the form of a radial heat flow and the
exterior space-time ~$V^{+}$~ was described by the
Reissner-Nordstrom-Vaidya metric which presents a spherically
symmetric electric field and a radial flow of unpolarized
radiation. Now in this paper we will consider the interior
space-time ~$V^{-}$~ is quasi-spherical Szekeres model and the
exterior space-time ~$V^{+}$~ is Reissner-Nordstrom-Vaidya metric
in the presence of electro-magnetic field.\\

\section{\normalsize\bf{Junction Conditions}}

Let ~$\Sigma$~ be time-like ~$3D$~ hypersurface which separates
two ~$4D$~ manifolds ~$V^{-}$~ and ~$V^{+}$~. The interior ~$4D$~
manifold  ~$V^{-}$~ is given by the quasi-spherical Szekeres
space-time with line element

\begin{equation}
V^{-}~~~:~~~ds^{2}_{-}=-dt^{2}+e^{2\alpha}dr^{2}+e^{2\beta}(dx^{2}
+dy^{2})
\end{equation}

where ~$\alpha$~ and ~$\beta$~ are functions of all space-time
co-ordinate variables.\\

Now for the transformation ~$$x=\cot \frac{\theta}{2}~\cos \phi,~
 y=\cot \frac{\theta}{2}~\sin \phi ,$$~ the metric ~$(1)$~ in
 ~$V^{-}$~becomes

\begin{equation}
ds^{2}_{-}=-dt^{2}+e^{2\alpha}dr^{2}+\frac{1}{4}e^{2\beta}cosec
\frac{\theta}{2}(d\theta^{2}+\sin^{2} \theta ~d\phi^{2})
\end{equation}

We assume that the source for the Einstein's field equations in
~$V^{-}$~ is given by

\begin{equation}
G_{\mu\nu}=T_{\mu\nu}+E_{\mu\nu}
\end{equation}

The stress-energy tensor of a non-viscous heat conducting fluid
has the expression

\begin{equation}
T_{\mu\nu}=(\rho+p)~u_{\mu}~u_{\nu} +p~g_{\mu\nu}
+q_{\mu}~u_{\nu}+q_{\nu}~u_{\mu}
\end{equation}

where ~$\rho,~p,~q_{\mu}$~ are the  fluid density, isotropic
pressure and heat flow vector respectively and ~$u_{\mu}$~ is the
four velocity. We take the heat flow vector ~$q_{\mu}$~ to be
orthogonal to the velocity vector ~$u^{\mu}$~ i.e.,
~$q_{\mu}~u^{\mu}=0 .$~ For co-moving co-ordinate system, we
choose ~$u^{\mu}=(1, 0, 0, 0)$~ and ~$q^{\mu}=(0, q_{r}, q_{x},
q_{y})$~ with ~$q_{i}=q_{i}(t, r, x, y),~~i\equiv(r, x, y)$.\\

The energy- momentum tensor ~$E_{\mu\nu}$~ of the electro-magnetic
field is given by

\begin{equation}
E_{\mu\nu}=\frac{1}{4\pi}\left[F^{\nu}_{\mu}~F_{\mu\nu}-\frac{1}{4}g_{\mu\nu}F^{\nu\delta}F_{\nu\delta}\right]
\end{equation}

where ~$F_{\mu\nu}$~ is the electro-magnetic field tensor
satisfying Maxwell's equations

\begin{equation}
F_{\mu\nu}=\phi_{\nu,\mu}-\phi_{\mu,\nu}
\end{equation}
and
\begin{equation}
F^{\mu\nu}_{~;~\nu}=4\pi J^{\mu}
\end{equation}

where ~$\phi_{\mu}$~ is the four-potential and ~$J_{\mu}$~ is the
four-current.\\

Since the charge is at rest in this system, so there will be no
magnetic field in this system. Thus, we may choose the
four-potential and four-current as

\begin{equation}
\phi_{\mu}=\left(\phi(t, r, x, y), 0, 0, 0\right)
\end{equation}
and
\begin{equation}
J^{\mu}=\sigma~u^{\mu}
\end{equation}

where ~$\sigma$~ is the charge density. From equation ~$(6)$~
using ~$(8)$~ we have the only non-zero component of the field
tensor as
\begin{equation}
F_{01}=-F_{10}=-\frac{\partial \phi}{\partial r}
\end{equation}

Also from equations ~$(7)$~ and ~$(9)$~ we obtain
\begin{equation}
\frac{\partial ^{2} \phi}{\partial r^{2}}-\alpha'~\frac{\partial
\phi}{\partial r}=4\pi\sigma e^{2\alpha}
\end{equation}
and
\begin{equation}
\frac{\partial ^{2} \phi}{\partial r
\partial t}-\dot{\alpha}~\frac{\partial
\phi}{\partial r}=0
\end{equation}

where `$~'~$' and `$~.~$' denote partial derivatives with respect
to ~$r$~ and ~$t$~ respectively. Integrating (12), we have

\begin{equation}
\frac{\partial \phi}{\partial r}=K(r, x, y)~e^{\alpha}
\end{equation}

and then from equation ~$(11)$~ one gets

\begin{equation}
K' =4\pi\sigma e^{\alpha}
\end{equation}

For the interior space-time model ~$(1)$~ with matter field
~$(4)$~ and electromagnetic field ~$(5)$, the Einstein's equation
~$(3)$~ can be explicitly written as the following :

\begin{equation}
2~\dot{\alpha}~\dot{\beta}+\dot{\beta}^{2}-e^{-2\beta}(\alpha^{2}_{x}+\alpha^{2}_{y}+\alpha_{xx}+\beta_{xx}+\alpha_{yy}
+\beta_{yy})+e^{-2\alpha}(2~\alpha'~\beta'-3~\beta^{'2}-2\beta'')=\rho-\frac{1}{8\pi}~K^{2}
\end{equation}

\begin{equation}
3~\dot{\beta}^{2}+2~\ddot{\beta}-e^{-2\alpha}~\beta^{'2}-(\beta_{xx}+\beta_{yy})e^{-2\beta}=-p-\frac{1}{8\pi}K^{2}
\end{equation}

\begin{equation}
\dot{\alpha}^{2}+\ddot{\alpha}+\dot{\alpha}\dot{\beta}+\dot{\beta}^{2}+
\ddot{\beta}+e^{-2\alpha}(\alpha'\beta'-
\beta'^{2}-\beta'')-e^{-2\beta} [\alpha_{x}^{2}+
\alpha_{xx}-\alpha_{x}\beta_{x}]-e^{-2\beta}\alpha_{y}\beta_{
y}=-p+\frac{1}{8\pi}K^{2}
\end{equation}

\begin{equation}
\dot{\alpha}^{2}+\ddot{\alpha}+\dot{\alpha}\dot{\beta}+\dot{\beta}^{2}+
\ddot{\beta}+e^{-2\alpha}(\alpha'\beta'-
\beta'^{2}-\beta'')-e^{-2\beta} [\alpha_{y}^{2}+
\alpha_{yy}-\alpha_{y}\beta_{y}] -e^{-2\beta}\alpha_{x}\beta_{ x}
=-p+\frac{1}{8\pi}K^{2}
\end{equation}

\begin{equation}
\alpha_{x}(-\alpha_{y}+\beta_{y})+\beta_{x}\alpha_{y}-\alpha_{xy}=0
\end{equation}
\begin{equation}
\alpha_{y}(-\alpha_{x}+\beta_{x})+\beta_{y}\alpha_{x}-\alpha_{xy}=0
\end{equation}

\begin{equation}
-\dot{\alpha}\beta'+\dot{\beta}\beta'+\dot{\beta}'=\frac{1}{2}~q_{r}~
e^{2\alpha}
\end{equation}

\begin{equation}
(\dot{\alpha}-\dot{\beta})\alpha_{x}+\dot{\alpha}_{x}+\dot{\beta}_{x}
=q_{x}~e^{2\beta}
\end{equation}

\begin{equation}
(\dot{\alpha}-\dot{\beta})\alpha_{y}+\dot{\alpha}_{y}+\dot{\beta}_{y}
=q_{y}~e^{2\beta}
\end{equation}

\begin{equation}
\alpha_{x}\beta'-\beta'_{x}=0
\end{equation}

\begin{equation}
\alpha_{y}\beta'-\beta'_{y}=0
\end{equation}

where subscript stands for partial derivatives with respect to
corresponding variables.\\

For quasi-spherical case i.e., for ~$\beta' \neq 0$~, we have the
solutions [17]

\begin{equation}
e^{\beta} = R(t,r)~e^{\nu(r, x, y)}
\end{equation}
and
\begin{equation}
e^{\alpha}=\frac{R'+R~\nu~'}{D(t, r)}
\end{equation}

where ~$R(t, r)$~ and ~$D(t, r)$~ will satisfy the differential
equations

\begin{equation}
2~R~\ddot{R}+(\dot{R}^{2}-D^{2})+\left(p+\frac{K^{2}}{8\pi}\right)R^{2}=f(r)
\end{equation}
and
\begin{equation}
R~\dot{D}=f(r)~e^{-2\alpha}
\end{equation}

with ~$f(r)$~ an arbitrary function of ~$r$~. The function ~$\nu$~
can be written in the form

\begin{equation}
e^{-\nu}=A(r)~(x^{2}+y^{2})+B_{1}(r)~x +B_{2}(r)~y+C(r)
\end{equation}

where ~$A, B_{1}, B_{2}$~ and ~$C$~ are arbitrary functions of
~$r$~ along with the restriction

\begin{equation}
(B_{1}^{2}+B_{2}^{2})-4~A~C=f(r)-1
\end{equation}

From equations ~(21), (22)~ and ~(23)~ we have the components of
heat flux vector as [17]

\begin{equation}
q_{r}=\frac{2}{R}~\dot{D}~e^{-\alpha}
\end{equation}

\begin{equation}
q_{x}=- \frac{\dot{D}}{D}~\alpha_{x}~e^{-\beta}
\end{equation}

\begin{equation}
q_{y}=- \frac{\dot{D}}{D}~\alpha_{y}~e^{-\beta}
\end{equation}

The exterior manifold ~$V^{+}$~ is described by the
Reissner-Nordstrom-Vaidya space-time having metric

\begin{equation}
V^{+} :~~~~ d
s^{2}_{+}=-\left[1-\frac{2M(v)}{z}+\frac{Q^{2}}{z^{2}}\right]d
v^{2}-2d v~d z+z^{2}(d \theta^{2}+\sin^{2}\theta~d \phi^{2})
\end{equation}

This space-time represents a spherically symmetric field with an
outgoing radial flux of unpolarized radiation and ~$v$~ is the
retarded time. \\

The surface of separation ~$\Sigma$~ between these two manifolds
~$V^{\pm}$~ is characterized by ~$r=r_{\Sigma}$~ and the intrinsic
metric on it (in the comoving co-ordinates of the interior
space-time $V^{-}$) is given by

$$ d s^{2}_{\Sigma}=-d \tau^{2}+R^{2}(\tau)(d \theta^{2}+\sin^{2}\theta~d
\phi^{2})$$

Now in the description of Santos, the Israel's
junction conditions are\\

(i)~The continuity of the line-element

\begin{equation}
(d s^{2}_{-})_{\Sigma}=(d s^{2}_{+})_{\Sigma}=d s^{2}_{\Sigma}
\end{equation}

where ~$(~~)_{\Sigma}$~ means the value of ~$(~~)$~ on ~$\Sigma$~. \\

(ii)~~The continuity of the extrinsic curvature over ~$\Sigma$~ :

\begin{equation}
[K_{ij}]=[K_{ij}^{+}]-[K_{ij}^{-}]=0  ,
\end{equation}

where the expressions of extrinsic curvature in terms of the unit
space-like normal vector to ~$\Sigma$~, ~$n_{\alpha}^{\pm}$~ are
given by [18],

\begin{equation}
K_{ij}^{\pm}=-n_{\sigma}^{\pm}\frac{\partial^{2}\chi^{\sigma}_{\pm}}{\partial\xi^{i}
\partial \xi^{j} }-n^{\pm}_{\sigma}\Gamma^{\sigma}_{\mu
\nu}\frac{\partial\chi^{\mu}_{\pm}}{\partial\xi^{i}}\frac{\partial\xi^{\nu}_{\pm}}
{\partial\xi^{j}}
\end{equation}

Here ~$\xi^{i}$~ refer to the intrinsic co-ordinates ~$(\tau, x,
y)$~ on ~$\Sigma$~ and ~$\chi^{\mu}_{\pm}~, ~\mu=0, 1, 2, 3$~
stand for the co-ordinates in ~$V^{\pm}$~. \\

The boundary ~$\Sigma$~ in terms of the co-ordinates of the
interior space-time ~$V^{-}$~ is given by

\begin{equation}
{\cal F}(r,t)=r-r_{\Sigma}=0~~~~~(r_{\Sigma}~,~~\text{a~constant})
\end{equation}

So one can write the normal vector ~$n^{-}_{\mu}$~ as
~$$n^{-}_{\mu}=(0, e^{\alpha}, 0, 0).$$~ Now comparing the metric
ansatz for ~$\Sigma$~ and ~$V^{-}$~ for ~$d r=0$~, the continuity
of the line element gives

\begin{equation}
\frac{d t}{d \tau}=1, ~~~~{\cal R}(\tau)=e^{\beta}~~~\text{
on}~~~~r=r_{\Sigma}
\end{equation}

Further, the components of the extrinsic curvature for the
interior space-time are

\begin{eqnarray}
\left.\begin{array}{llll}
 K^{-}_{\tau\tau}=0,~~~K^{-}_{\theta\theta}=cosec^{2}\theta~K^{-}_{\phi\phi}
 =\left[\frac{1}{4}\beta^{'2}~e^{2\beta-\alpha}~cosec^{4}(\frac{\theta}{2})\right]_{\Sigma}\\\\
 \text{and}~~~K^{-}_{ij}=0,~~~i \neq j
\end{array}\right\}
\end{eqnarray}

On the other hand, in terms of the co-ordinates of the
Reissner-Nordstrom-Vaidya space-time the surface ~$\Sigma$~ can be
characterized by

\begin{equation}
{\cal F}(z, v)=z-z_{\Sigma}(v)=0
\end{equation}

Hence, as before the unit normal vector to ~$\Sigma$~ is given by
\begin{equation}
n^{+}_{\mu}=\left(2\frac{d z_{\Sigma}}{d
v}+1-\frac{2M(v)}{z_{\Sigma}}+\frac{Q^{2}}{z_{\Sigma}^{2}}\right)^{-\frac{1}{2}}\left(-\frac{d
z_{\Sigma}}{d v}, 1, 0, 0\right)
\end{equation}

and the components of the extrinsic curvature for the exterior
space-time are

\begin{equation}
\left.\begin{array}{llll}
 K^{+}_{\tau\tau}=\left[\frac{\ddot{v}}{\dot{v}}-\dot{v}\left(\frac{M(v)}{z^{2}}-\frac{Q^{2}}{z^{3}}\right)\right]_{\Sigma}~,\\\\
K^{+}_{\theta\theta}=~cosec^{2}~
\theta~K^{+}_{\phi\phi}=\left[\dot{v}\left(1-\frac{2M(v)}{z}+\frac{Q^{2}}{z^{2}}\right)z+z\dot{z}\right]_{\Sigma}~,\\\\
K^{+}_{ij}=0,~~~i \neq j
\end{array}\right\}
\end{equation}

(here ~`$~.~$'~ stands for the differentiation with respect to
~`$~\tau~$'~).\\

Now, the continuity of the metric ansatz for ~$\Sigma$~ and
~$V^{+}$~ over ~$z=z_{\Sigma}(v)$~ is given by

\begin{eqnarray}
\left.\begin{array}{llll}
 z_{\Sigma}(v)=R(\tau)~,\\\\
 \left[2\frac{d z}{d
v}+1-\frac{2M(v)}{z}+\frac{Q^{2}}{z^{2}}\right]_{\Sigma}=\left(\frac{1}{\dot{v}^{2}}\right)_{\Sigma}
\end{array}\right\}
\end{eqnarray}

Also the continuity of the components of the extrinsic curvature
across ~$\Sigma$~ give
\begin{equation}
p=q_{r}~e^{\alpha}+\frac{f(r)-1}{R^{2}}+\frac{4}{R^{2}}~sin^{4}(\theta/2)e^{-2\nu}
-\frac{Q^{2}}{R^{4}}-\frac{K^{2}}{8\pi}
\end{equation}
and
\begin{equation}
M(v)=\frac{R}{2}+\frac{Q^{2}}{2~R}-\frac{D^{2}~R}{2}+\frac{R~\dot{R}^{2}}{2}
\end{equation}

which can be termed as the total energy entrapped inside the
surface ~$\Sigma$~. We note that on the boundary, the pressure
can not be vanish in general. Thus for quasi-spherical shear-free
distribution of a collapsing charged fluid undergoing dissipation
in the form of heat flow, the isotropic pressure on the surface of
discontinuity can not be zero. In particular, if fluid stops
dissipation, ~$q_{r}=0$, the pressure can not vanish at the
boundary, but in this case radiation can not exist and the
exterior space-time ~$V^{+}$~ is the Reissner-Nordstrom-Vaidya
space-time. But in absence of isotropic pressure there may still
be radiation on the boundary and the exterior space-time ~$V^{+}$~
will still be Reissner-Nordstrom-Vaidya space-time.\\

The total luminosity for an observer rest at infinity is [7]

\begin{equation}
L_{\infty}=-\left(\frac{dm}{dv}\right)_{\Sigma}
\end{equation}

If we now consider an observer on the boundary ~$\Sigma$~ then
the luminosity for that observer is

\begin{equation}
L_{\Sigma}=-\left[\left(\frac{dv}{d\tau}\right)^{2}~\frac{dm}{dv}\right]_{\Sigma}
\end{equation}

Now the boundary redshift of the radiation emitted by a star can
be written as

\begin{equation}
Z_{\Sigma}=\sqrt{\frac{L_{\Sigma}}{L_{\infty}}}~-1=\frac{dv}{d\tau}-1=\frac{1}{\dot{R}+D}-1
\end{equation}

Hence the luminosity measured by an observer at rest at infinity
is reduced by the redshift in comparison to the luminosity
observed on the surface of collapsing body. Also when ~$\dot{R}+D
=0 $ then the boundary redshift attains unlimited value (i.e.,
~$Z_{\Sigma}\rightarrow \infty$~).\\

\section{\normalsize\bf{Apparent Horizon and Its Time of Formation }}
The first integral of equation ~$(28)$~ is given by

\begin{equation}
\dot{R}^{2}=f(r)+\frac{F(r)}{R}-\frac{K^{2}R^{2}}{24\pi}+\frac{1}{R}~\int(D^{2}-pR^{2})dR
\end{equation}

As ~$D$~ is regular and ~$p$~ blows up at the singularity [19], so
let us assume

\begin{equation}
D=\lambda(r)R ,~~~p=\frac{f(r)}{R^{2}}
\end{equation}
The evolution equation ~$(51)$~ satisfies to
\begin{equation}
\dot{R}^{2}=\frac{F(r)}{R}-\frac{K^{2}R^{2}}{24\pi}+\frac{\lambda^{2}}{3}R^{2}
\end{equation}

Further, the time of formation of apparent horizon ~$t_{ah}(r)$~
is given by
\begin{equation}
\dot{R}^{2}(t_{ah}(r), r)=1+f(r)
\end{equation}
Hence for equation (53) we have a cubic equation in ~$R$~ as
\begin{equation}
\left(\frac{\lambda^{2}}{3}-\frac{K^{2}}{24\pi}\right)R^{3}(t_{ah}(r),r)-(1+f(r))R(t_{ah}(r),r)+F(r)=0
\end{equation}
Thus apparent horizon exists if this cubic equation in ~$R$~ has
at least one positive root. We shall show below different
conditions for existence of positive roots.\\

I.~~For ~$\frac{\lambda^{2}}{3}-\frac{K^{2}}{24\pi} > 0$~ and
~$F(r) <
\frac{2(1+f)^{\frac{3}{2}}}{\left(\frac{\lambda^{2}}{3}-\frac{K^{2}}{24\pi}\right)^{\frac{1}{2}}}$~,
there are two positive roots of equation (55) and hence there are
two apparent horizons, given by

\begin{equation}
R_{c}(r)=2
\sqrt{\frac{(1+f)^{3}}{(\frac{\lambda^{2}}{3}-\frac{K^{2}}{24\pi})}}
\cos\left[\frac{1}{3}\cos^{-1}\left(-\frac{1}{2}F(r)\sqrt{\frac{\lambda^{2}}{3}-\frac{K^{2}}{24\pi}}\right)\right]
\end{equation}

\begin{equation}
R_{b}(r)=2
\sqrt{\frac{(1+f)^{3}}{(\frac{\lambda^{2}}{3}-\frac{K^{2}}{24\pi})}}
\cos\left[\frac{4\pi}{3}+\frac{1}{3}\cos^{-1}\left(-\frac{1}{2}F(r)\sqrt{\frac{\lambda^{2}}{3}-\frac{K^{2}}{24\pi}}\right)\right]
\end{equation}

which are termed as cosmological and black hole horizons.\\

II.~~When
~$F(r)=\frac{2(1+f)^{\frac{3}{2}}}{\left(\frac{\lambda^{2}}{3}-\frac{K^{2}}{24\pi}\right)^{\frac{1}{2}}}~,~~~~
(\lambda > \frac{K}{2\sqrt{2\pi}})$~ there is only one positive
root, which corresponds to a single apparent horizon
\begin{equation}
R_{bc}(r)=
\sqrt{\frac{(1+f)^{3}}{(\frac{\lambda^{2}}{3}-\frac{K^{2}}{24\pi})}}
\end{equation}

III.~~For
~$F(r)>\frac{2(1+f)^{\frac{3}{2}}}{(\frac{\lambda^{2}}{3}-\frac{K^{2}}{24\pi})^{\frac{1}{2}}},~~~~
(\lambda > \frac{K}{2\sqrt{2\pi}})$~ there are no positive roots
and consequently there are no apparent horizons.\\

IV.~~If ~$\lambda < \frac{K}{2\sqrt{2\pi}}$~ then also one
positive root exists to have a unique apparent horizon. A detailed
study of apparent horizons in quasi-spherical gravitational
collapse with a non-zero ~$\Lambda$-term can be found in
[20].\\

Moreover, solving the dynamical equation (53) we obtain

\begin{equation}
R^{\frac{3}{2}}=\frac{2\sqrt{6\pi
F}}{\sqrt{8\pi\lambda^{2}-K^{2}}}~\sinh\left[\frac{\sqrt{8\pi\lambda^{2}-K^{2}}}{4\sqrt{\frac{2\pi}{3}}}(t-t_{s}(r))\right]
\end{equation}

where ~$t_{s}(r)$~ is the time of formation of singularity of a
particular shell at co-ordinate distance ~`$r$' ~ (i.e.,~ $R=0$
at $t_{s}(r)=0)$. Hence the time difference between the formation
of apparent horizon and the singularity formation is given by

\begin{equation}
t_{ah}(r)-t_{s}(r)=\frac{4\sqrt{\frac{2\pi}{3}}}{\sqrt{8\pi\lambda^{2}-K^{2}}}~\sinh^{-1}\left[\frac{\sqrt{8\pi\lambda^{2}-K^{2}}}
{2\sqrt{6 \pi F}}~R^{\frac{3}{2}}_{ah}\right]
\end{equation}

(One may note that in solving the differential equation (53) it is
assumed that ~$ \lambda > \frac{K}{2\sqrt{2\pi}} $~. However if
~$\lambda < \frac{K}{2\sqrt{2\pi}}$~ then ~`sinh'~ is to be
replaced by ~`sin'~ in the solution (59)). Thus it is to be
observed that heat flow and electromagnetic field modify the
formation of horizon and also the time difference between the
formation of
apparent horizon and singularity.\\

Further, from equation ~$(47)$~ we note that there is dissipation
of total energy entrapped inside the surface ~$\Sigma$~ due to the
heat flow while the electromagnetic field favours entrapped energy
within the surface ~$\Sigma$~. Finally, the contribution to the
isotropic pressure due to the heat flow and electromagnetic field
is in reverse order (or in the same order) provided ~$D$~ is
positive (or negative). Hence for increasing the electromagnetic
field tries to diminish the pressure on the boundary which may be
balanced by the heat flux term, while for decreasing ~$D$~ both of
them try to reduce the pressure. So we conclude that
electromagnetic field favours formation of naked singularity while
heat flux may or may not be favourable for formation of black
hole.\\\\

{\bf Acknowledgement:}\\

One of the authors (UD) is thankful to CSIR,
Govt. of India for providing research project grant (No. 25(0153)/06/EMR-II).\\

{\bf References:}\\
\\
$[1]$  J. R. Oppenhiemer and H. Snyder, {\it Phys. Rev.} {\bf 56} 455 (1939).\\
$[2]$  C. W. Misner and D. Sharp, {\it Phys. Rev.} {\bf 136} B571
(1964).\\
$[3]$  P. C. Vaidya, {\it Proc. Indian Acad. Sci. A} {\bf 33} 264
(1951).\\
$[4]$  N. O. Santos, {\it Phys. Lett. A} {\bf 106} 296 (1984); {\it Mon. Not. R. Astr. Soc.} {\bf 216} 403 (1985).\\
$[5]$  R. W. Lindquist, R. A. Schwartz and C. W. Misner, {\it
Phys. Rev.} {\bf 137} B1364 (1965).\\
$[6]$  W. Israel, {\it Nouvo Cimento} {\bf 44B} 1 (1966); {\bf 48B} 463  (1966); {\it Phys. Lett.} {\bf 24A} 184 (1967).\\
$[7]$  A. K. G. de Oliveira, N. O. Santos and C. A. Kolassis, {\it
Mon. Not. R. Astr. Soc.} {\bf 216} 1001
(1985).\\
$[8]$  A. K. G. de Oliveira, J. A. de F. Pacheco and N. O. Santos,
{\it Mon. Not. R. Astr. Soc.} {\bf 220} 405
(1986).\\
$[9]$  A. K. G. de Oliveira and N. O. Santos, {\it Astrophys. J.}
{\bf 312} 640 (1987).\\
$[10]$  A. K. G. de Oliveira, C. A. Kolassis and N. O. Santos,
{\it Mon. Not. R. Astr. Soc.} {\bf 231} 1011 (1988).\\
$[11]$ C. W. Misner, {\it Phys. Rev.} {\bf 137B} 1360 (1965).\\
$[12]$ C. W. Misner and D. H. Sharp, {\it Phys. Lett.} {\bf 15} 279 (1965).\\
$[13]$ K. Lake and C. Hellaby, {\it Phys. Rev. D} {\bf 24} 3019 (1981).\\
$[14]$ K. Lake, {\it Phys. Rev. D} {\bf 26} 518 (1982).\\
$[15]$  S. G. Ghosh and D. W. Deshkar, {\it Int. J. Mod. Phys. D}
{\bf 12} 317 (2003).\\
$[16]$  S. G. Ghosh and D. W. Deshkar, {\it Gravitation and
Cosmology} {\bf 6} 1 (2000).\\
$[17]$  U. Debnath, S. Nath and S. Chakraborty,
{\it Gen. Rel. Grav.} {\bf 37} 215 (2005).\\
$[18]$ L. P. Eisenhart, {\it Riemannian Geometry}, p. 146 (1949),
Princeton.\\
$[19]$ S. Chakraborty, S. Chakraborty and U. Debnath,
{\it Int. J. Mod. Phys. D} {\bf 14} 1707 (2005).\\
$[20]$ U. Debnath, S. Nath and S. Chakraborty,
{\it  Mon. Not. R. Astr. Soc.} {\bf 369} 1961 (2006).\\

\end{document}